\documentclass{iau_FM}
\usepackage{graphicx}

\title[A Radio Polarimetric Study of a Sample of Seyfert and starburst galaxies] 
{Understanding the Origin of Radio Outflows in Seyfert Galaxies using Radio Polarimetry }

\author[Sebastian et al.]   
{Biny Sebastian,$^{1}$\thanks{E-mail: biny@ncra.tifr.res.in}
P. Kharb,$^{1}$
C. P. O' Dea,$^{2,3}$
J. F. Gallimore,$^{4}$
S. A. Baum,$^{3,5}$
and E. J. M. Colbert$^{6,7}$
}

\affiliation{$^{1}$ National Centre for Radio Astrophysics (NCRA) - Tata Institute of Fundamental Research (TIFR), S. P. Pune University Campus, Ganeshkhind, Pune 411007, India\\
$^{2}$ School of Physics \& Astronomy, Rochester Institute of Technology, Rochester, NY 14623\\
$^{3}$ Physics and Astronomy, University of Manitoba, Winnipeg, Canada\\
$^{4}$ Department of Physics and Astronomy, Bucknell 
University, Lewisburg, PA 17837\\
$^{5}$ Carlson Center of Imaging Science, Rochester Institute of Technology, Rochester, NY 14623\\
$^{6}$ Hume Center for National Security and Technology, 900 N. Glebe Rd, Arlington, VA 22203\\
$^{7}$ U.S. Army Research Laboratory Adelphi, MD, USA \\
}

\pubyear{2019}
\jname{Astronomy in Focus, Volume 1} 
\editors{Piero Benvenuti, ed.}
\begin{document}

\maketitle

\begin{abstract}
The role of starburst winds versus active galactic nuclei (AGN) jets/winds in the formation of the kiloparsec scale radio emission seen in Seyferts is not yet well understood. In order to be able to disentangle the role of various components, we have observed a sample of Seyfert galaxies exhibiting kpc-scale radio emission suggesting outflows, along with a comparison sample of starburst galaxies, with the EVLA B-array in polarimetric mode at 1.4 GHz and 5~GHz. Polarization is clearly detected in three Seyfert galaxies and one starburst galaxy. The Seyfert galaxy NGC\,2639, shows highly polarized secondary radio lobes, not observed before, which are aligned perpendicular to the known pair of radio lobes. The additional pair of lobes represent an older epoch of emission. A multi-epoch multi-frequency study of the starburst-Seyfert composite galaxy NGC\,3079, reveals that the jet together with the starburst superwind and the galactic magnetic fields might be responsible for the well-known 8-shaped radio lobes observed in this galaxy. We find that many of the Seyfert galaxies in our sample show bubble-shaped lobes, which are absent in the starburst galaxies that do not host an AGN.
\end{abstract}

\firstsection 
\section{Introduction}
It is known that many active galactic nuclei (AGN) like radio galaxies and quasars host very powerful radio jets, which can extend up to several hundreds of kpc or sometimes even up to a few Mpcs. However, rarely do Seyfert galaxies host such powerful jets, although several high angular resolution studies have revealed that many of them possessed double or triple components similar to radio galaxies (Ulvestad et al. 1981), which are sometimes distorted into S-shapes, or are loop-like, etc. Lower angular resolution studies on the other hand, have shown that most of these Seyfert galaxies possess radio emission that sometimes extends up to several kpcs. However, the origin of the kiloparsec-scaled radio structure (KSR) emission is debated in the literature. For example, Baum et al. 1993 have concluded that the KSR was a consequence of starburst winds, since most of the emission was aligned along the minor axis of the host galaxy. Colbert et al. 1996, on the other hand carried out a similar study of a sample of Seyfert galaxies but with a comparison sample of starburst galaxies. They found that while the radio powers are comparable, the morphologies of the kilo-parsec scale emission are different for a sample of starburst galaxies which are more diffuse versus that of the Seyfert galaxies which are more lobe-like and oriented at skewed angles to the minor axis.
Gallimore et al. 2006 have carried out a more complete survey of a sample of 43 Seyfert galaxies, and they find that at least 44\% of these galaxies host kilo-parsec scale radio emission. While they favored an AGN driven origin for these outflows, they did not rule out the role of starburst superwinds. 
It was demonstraed by Irwin et al. 2017 that polarization can prove to be an effective tool in distinguishing the AGN-jet related emission from that of the galactic disk radio continuum emission. 
The radio lobes, which were embedded in the disk emission in total intensity images, were revealed in the polarization intensity image due to the higher degrees of linear polarization of the lobes compared to the rest of the host galaxy disk emission. In this paper, we present results from an ongoing work, where we are trying to understand if polarization can be used to distinguish outflows which have a jet related origin, versus those with a starburst wind driven origin. We are trying to address questions like (i) whether we will find higher degrees of polarization in a sample of Seyfert galaxies compared to starburst galaxies, (ii) if we will find signatures of ordered magnetic fields in Seyfert galaxy outflows more often than in starburst galaxy outflows, and (iii) whether the alignment of the magnetic fields in Seyfert galaxies will be different from that expected to be generated by host galaxy dynamo mechanism as observed in typical starburst galaxies. We have chosen a sample of starburst galaxies along with a sample of Seyfert galaxies, which we describe below.

\section{Sample and Observations}
We chose 10 Seyferts from the CfA (Huchra \& Burg, 1992) and 12 $\mu$m sample (Rush et al. 1993), which possessed lobe to lobe extents $\geq 20''$ from the study of Gallimore et al. 2006. We chose a comparison sample of 8 edge-on starburst galaxies from Colbert et al. 1996 and Dahlem et al. 1998. 
The sample was observed using the EVLA in the B-array configuration. Four of the starburst sources were observed using L-band (1.5 GHz), and the rest of the sources were observed using the C-band (5.5 GHz). Additional data are being obtained at X-band (10.0 GHz) using the D-array configuration under the project code (19B-198), to decrease the effect of wavelength dependant depolarization effects dominant at lower wavelengths. 
\section{Results}
We present the results of the detailed analysis of two Seyfert galaxies where we detected polarization, in Sections 3.1 and 3.2, and some early results from the larger sample in Section 3.3. 

\subsection{Filamentary lobes in the starburst-Seyfert Composite Galaxy NGC\,3079}
\begin{figure*}
\centerline{
\includegraphics[width=6.0cm]{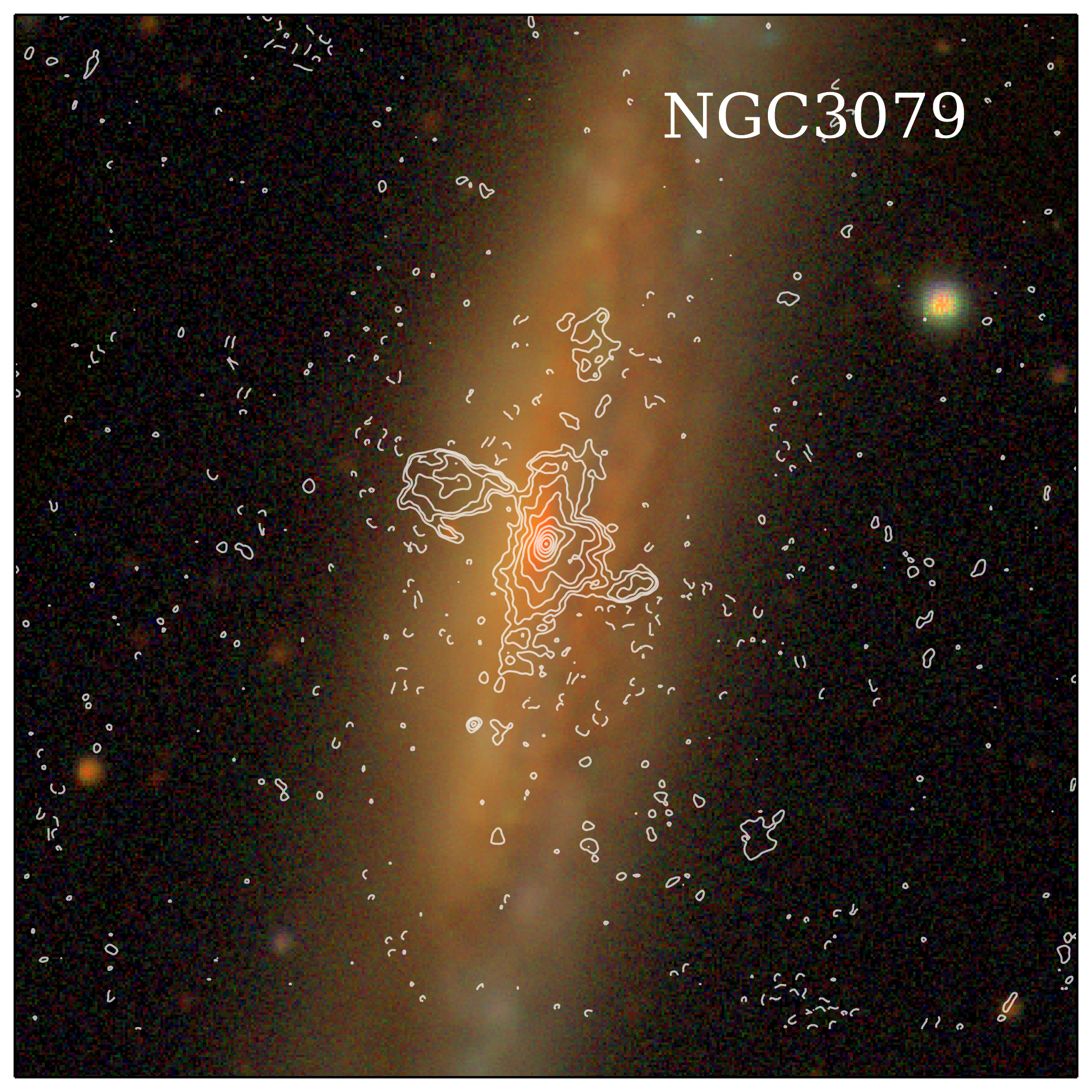}
\includegraphics[width=6.7cm,trim=40 160 0 140]{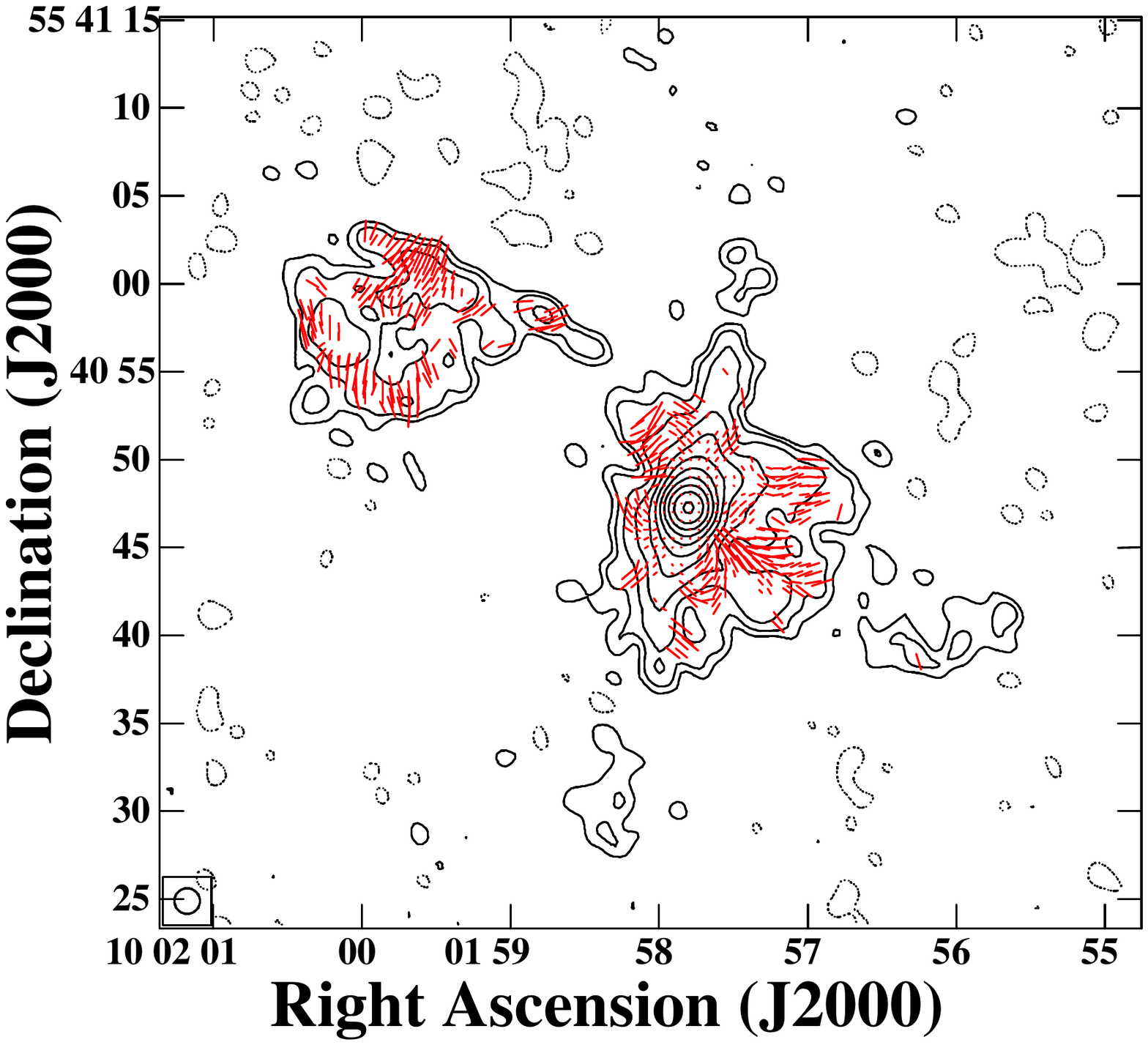}}
\caption{\scriptsize{(Left) 1.5 GHz EVLA A-array image contours with levels 145 $\times$ (-1, 1, 2, 4, 8, 16, 64, 32, 64, 128) $\mu$Jy beam$^{-1}$ overlaid on the optical SDSS {\it gri} color composite image, 
and (right) 5 GHz VLA B-array image with contour levels 0.80 $\times$ (-0.085, 0.085, 0.17, 0.35, 0.70, 1.40, 2.80, 5.60, 11.25, 22.50, 45, 90) mJy beam$^{-1}$ of NGC\,3079 with polarization electric vectors whose length is proportional to the fractional polarization (1 arcsec length corresponds to 40\% fractional polarization), superimposed in red.}}
\label{fig2}
\end{figure*}

NGC\,3079 is a Seyfert galaxy with a prominent starburst component (Dahlem et al. 1998). Duric \& Seaquist 1988 identified the ``eight'' shaped morphology of the lobes in NGC\,3079. They suggested that these lobes which are aligned along the minor axis of the galaxy are generated as a result of winds that are either powered by a starburst or by the AGN. On the other hand, Irwin \& Seaquist 1988 who studied the VLBI jet argued that the jet can essentially power the entire lobes. X-ray and emission line imaging showed the presence of superwinds present in NGC\,3079. We have carried out a detailed study of NGC\,3079 using multi-frequency legacy VLA observations which complemented our current EVLA observations (Sebastian et al. 2019a). Our sensitive high resolution images revealed the complex filamentary morphology of the lobes. Morphologically, the radio lobes in NGC\,3079 do not resemble those seen in powerful radio galaxies, which either show a hotspot and a backflow (in the case of FR\,II type radio galaxies) or diffuse plumes or tails which become more diffuse and uncollimated with distance from the core (in the case of FR\,I type radio galaxies). 
We investigated other possible mechanisms that might be influencing the formation of the lobes with bubble like morphology. For example, shocks are a feasible mechanism which can explain many of the observed features of the ``ring'' or loop-like like structure that are observed inside the northern lobe. Shocks, can give rise to synchrotron emission via diffusive shock acceleration (Blandford \& Eichler 1987). Shock compression at the boundaries of the expanding bubble and the ``ring'' can also lead to the amplification of magnetic fields, which ultimately result in the high fractions of polarization ($\sim$33\%) along the ``ring'' (see Figure~\ref{fig2}). The flatter spectral indices observed along the edges of the ring can also be explained as a result of re-acceleration of relativistic particles along the edges. 

We also investigated how well the radio emission correlated with the thermal components of the superwind using the Chandra ACIS-S X-ray observations and HST WFPC2 H$\alpha$+ [NII] line emission data. We noticed that the thermal emission peaked to the south of the region where the radio emission peaked. Moreover, X-ray and the emission line filaments did not appear very well correlated spatially, as can be seen in Figure 9 in Sebastian et al. 2019a. Adebahr et al. 2013 found that the thermal emission traced by emission lines is correlated with the radio continuum emission in M82, a prototypical starburst galaxy hosting superwinds, which is suggestive of frozen-in magnetic fields. Brandenburg et al. 1995 suggested that disk material and frozen in magnetic fields are expected to be entrained along with winds. However, the lack of a correlation between radio and X-ray/emission line filaments in NGC\,3079, argue against this scenario. Another interesting aspect about this source is that the filamentary structure in the northern lobes shows a rotation measure inversion (see Cecil et al. 2001 \& Sebastian et al. 2019a). It is hard to explain such an inversion only by using shock acceleration. Organized magnetic fields on the scales of the kpc-scale ring are required to explain such a geometry.

Cecil et al. 2001 have suggested that the RM inversion seen in NGC\,3079 can be explained as a result of loop-like and twisted magnetic field structures similar to solar prominences. The dome-like structures that are observed in NGC\,3079 and active regions can be explained to have been generated as a result of the $\alpha^2$ dynamo mechanism in the galaxy. The synchrotron emission seen in the loop, however is higher than that can be produced by the winding at the base of the filaments. The radio-excess seen compared to the standard radio-FIR correlation (Sebastian et al. 2019a), along with the presence of the mildly relativistic jet (Irwin \& Seaquist 1988; Trotter et al. 1998; Sawada-Satoh et al. 2000) together points to a possibility that the relativistic plasma is supplied by the jet which gets frustrated by interaction with the surrounding medium on shorter length scales. It may be possible that the disrupted jet material gets transported out along with the superwinds, near the centre, where the thermal pressure is dominant (the region where X-ray or emission line emission is dominant) and later along the force-free magnetic fields, where the magnetic pressure and consequently the synchrotron emission is dominant. We therefore concluded that the peculiar lobe morphology observed in NGC\,3079 was caused by an interplay of the relativistic jet, the galactic dynamo and the starburst superwinds in the galaxy.

\subsection{Discovery of an Additional Pair of Radio Lobes in NGC\,2639}
With our polarization sensitive observations using the EVLA we were able to discover a pair of outer lobes in the north-south direction in addition to the already known pair of lobes which was aligned in the east-west direction (Sebastian et al. 2019b). The north-south lobes were highly polarized with fractional polarizations $\sim$21$\pm$4\%. 
There are several possible explanations like environmental influence, slow precession of the jets, the presence of a binary black hole at the core, and episodic activity.
However, it is unlikely that precession is the leading cause because of the lack of a bridge like feature connecting both pairs of lobes, similar to that seen in typical X/S-shaped galaxies (Lal et al. 2019). Also the position angle of the host galaxy minor axis is located almost midway between the north-south and east-west lobes, disfavoring any pressure gradients due to the galaxy itself as a reason for the misalignment. The probability of the existence of binary black holes both launching kpc-scale jets simultaneously or within the gap of the typical synchrotron loss time-scale is low and hence may not be the reason for the double pair of lobes. We favor the scenario where the origin of the two pair of lobes is due to the switching off and restarting of the central engine. The optical host galaxy shows a rather settled morphology. Hence, such a restarting might have been induced probably by a minor merger which changed the direction of the spin of the central black hole, but did not disturb the host galaxy morphology. 
The spectral age of the secondary lobes was $\sim$16 Myr. There have therefore been at least two AGN jet episodes in the past 16 Myrs.
\subsection{Starburst versus Seyfert Sample Properties}
We detect polarization in four out of the sixteen sources in our combined sample which includes three Seyfert galaxies, viz., NGC\,3079 (Sebastian et al. 2019a), NGC\,2639 (Sebastian et al 2019b) and NGC\,5506 (Kharb et al. 2020, in prep.) and one starburst galaxy, NGC\,253 (Sebastian et al. 2020, in prep.) 
We need to explain the following morphological differences that are observed in the sample of starbursts versus Seyferts.
\begin{itemize}
    \item Edge-brightened bubble-like radio lobes similar to those found in NGC\,3079, which make them distinct from the typical large-scale jets/lobes observed in radio galaxies are common in Seyfert galaxies but are absent in starbursts, without an AGN. These lobes are frequently roughly aligned with the minor axis, especially on several kpc-scales, but not always. We are exploring the role of the relativistic AGN jet in conjunction with the galactic dynamo in the formation of these lobes in the Seyfert sample.
    \item We find that the total intensity distribution of the lobes are rotationally symmetric on both sides, which is indicative of the role played by the central engine rather than being purely environment related.
    \item We find that many of the Seyfert galaxies (5/9) show multiple radio structures which are misaligned to each other on varying scales. However, in starburst galaxies, the distribution of radio continuum emission at different scales is often continuous and follows the star forming disk morphology or are aligned along the minor axis. We will explore if these misaligned structures represent multiple epochs of activity in our upcoming paper.
    \item For the sources in which we detected polarization, the Seyfert galaxies seem to show higher degrees of polarization compared to the starburst galaxies, at the scales probed by our observations. This is a tentative result and needs to be confirmed with deeper observations and larger samples.
\end{itemize}
\vspace{-10pt}
\section{Summary}
We are trying to use polarization to distinguish between starburst/ AGN driven radiative winds versus the jet driven outflows in Seyfert galaxies. We have detected polarization in three Seyfert galaxies and one starburst galaxy. The peculiar morphology of the lobes in NGC\,3079, which is a complex starburst galaxy also hosting a Seyfert nuclei, is a result of the interplay of the magnetised jet, the starburst wind and the galactic magnetic fields (Sebastian et al 2019a). We have discovered a new pair of radio lobes in the north-south direction in the Seyfert galaxy NGC\,2639; these lobes are aligned in a direction almost perpendicular to the previously known east-west lobes. The radio bubble-like lobes which are absent in starburst galaxies without an AGN are very common in Seyfert galaxies, and many a times not aligned with the minor axis, indicating the role of AGN in the formation of these lobes. We expect that our new observations using the EVLA at 10~GHz will enable us to nail down the reasons behind their origin and the differences between starburst and Seyfert galaxy radio emission.

\section{Acknowledgements}
The author is thankful to the organisers of the IAU Symposium 356 for an oral presentation and financial support to attend the conference. 
Baum and O'Dea are grateful to the Natural Sciences and Engineering Research Council of Canada (NSERC) for support. 



\end{document}